\documentstyle[12pt]{article}
\begin{document}
\makeatletter
\@addtoreset{equation}{section}
\makeatother
\renewcommand{\theequation}{\thesection.\arabic{equation}}
\baselineskip 15pt

\title{\bf Contextuality, Nonlocality and Counterfactual Arguments }

\author{ GianCarlo Ghirardi\footnote{e-mail: ghirardi@ts.infn.it}\\
{\small Department of Theoretical Physics of the University of Trieste, and}\\
{\small the Abdus Salam International Centre for Theoretical Physics,
Trieste, Italy, and}\\{\small the Istituto Nazionale di Fisica Nucleare.} \\ and\\
Karl Wienand\footnote{e-mail: Karl Wienand@gmail.com}\\
{\small Department of Theoretical Physics, University of
  Trieste, Italy}}
\date{}
\maketitle

\begin{abstract}
{\bf Abstract} In this paper, following an elementary line of thought which somewhat differs from the usual one, we prove once more that any deterministic theory predictively equivalent to quantum mechanics unavoidably exhibits a  contextual character. The purpose of adopting this perspective is that of paving the way for a critical analysis of the use of counterfactual arguments when dealing with nonlocal physical processes.\end{abstract}
\begin{quote}
\noindent {\it One of us (G.C.G) dedicates  this paper to  the colleague and  friend Pekka Lahti with whom he had many stimulating scientific and human interactions.}
\end{quote}

\section {Introduction}

In this paper very few original results are presented but the crucial problems of quantum nonlocality and of the unavoidable contextuality of any deterministic hidden variable theory equivalent to it are reconsidered by adopting a perspective which  differs from the one  which is usually found in the literature. The  reasons for presenting the arguments in this way stem from the desire to discuss some of the fundamental problems of our best theory in  a simple and straightforward way which might turn out to be illuminating; the advantages of the approach derive from the fact that it will make easily understandable the questions we will raise concerning the use of counterfactual arguments within a nonlocal context. We call attention to the fact that the choice of dealing with  deterministic hidden variable models is due to the ensuing extreme simplicity of the argument; the reader will meet no difficulty in realizing that analogous problems with the use of counterfactual reasoning emerge in general for any theory in which nonlocal effects occur.

\section{A first example: the case of two spin-1/2 particles}

To warm up, let us consider a system of two far away spin $\frac{1}{2}$ particles and let us denote as ${\bf A}^{(1)}$ and ${\bf B}^{(2)}$ the observables $\sigma^{(1)}\cdot{\bf a}$ and $\sigma^{(2)}\cdot{\bf b}$, respectively, with obvious meaning of the symbols. We suppose that the composite system is in a pure quantum state associated to the statevector $\vert\psi\rangle$. We will be interested in two different experimental situations: in the first, only one particle (let us say the i-th, i=1,2) is subjected to a measurement process of the observable ${\bf X}^{(i)}=\sigma^{i}\cdot {\bf x}$, while in the second, two measurements, one for each particle, are performed,  the corresponding observables being the pair $({\bf X}^{(i)},{\bf Y}^{(j)}), \hspace {0.2 cm}( i\neq j).$

As already anticipated, instead of resorting to the quantum standard approach, we will deal with an hypothetical deterministic hidden variable theory predictively equivalent to it  in which the complete specification of the state is given by assigning the value $\lambda$ of the hidden variables. When $\lambda$ is specified, all conceivable outcomes of prospective measurements on the physical system, in particular those of any conceivable single observable as well as those of the pairs corresponding to correlation experiments, take  one of the two possible values $\{ +1,- 1\}$. Equivalently, the assignement of $\lambda$ determines uniquely the probabilities\footnote{We have attached a star to the single measurement probabilities to remind the reader that they refer to cases in which the partner particle is not subjected to a measurement.} $P_{\lambda}({\bf X}^{(1)}=\alpha\vert *),\;\;\;P_{\lambda}(*\vert{\bf Y}^{(2)} = \beta),\;\;\;P_{\lambda}({\bf X}^{(1)}=\alpha,{\bf Y}^{(2)}=\beta)$ of the single and joint measurements outcomes. These probabilities can take only the values $\{0,1\}$ since the theory is deterministic. We will denote as $\Lambda$ the set onto which the hidden variables run, and as $\rho_{\psi}(\lambda)$ the distribution of the hidden variables associated to the situation which, in the corresponding quantum case, would be characterized by the statevector $\vert\psi\rangle$.

Let us consider now  the observable ${\bf A}^{(1)}$ of the first particle, and let us denote as $({\bf a}_{\lambda}=i)$ the value it takes for the considered $\lambda$. We further consider a measurement of the pair $({\bf A}^{(1)},{\bf B}^{(2)})$ and we denote as $({\bf a}_{\lambda}=i,{\bf b}_{\lambda}=j)$ the pair of outcomes obtained by subjecting both particles to the indicated measurements.

We make now  an assumption which will play an important role for the subsequent derivation:
\begin{quote}
{\bf Assumption L}: {\it for any ${\bf a}$ and $\lambda\in\Lambda$, the outcome $({\bf a}_{\lambda}=i)$ of the measurement performed on the first particle coincides with the outcome for the same observable ${\bf A}^{(1)}$ when two measurements are performed on the two particles, the first one referring to the same direction ${\bf a}$ and the second to an arbitrary direction ${\bf y}$, independently of the choice for ${\bf y}$.  In brief, we assume  that if, e.g., $({\bf a}_{\lambda}=+1)$, then one has $({\bf a}_{\lambda}=+1,{\bf y}_{\lambda}=j)$ for the double measurement, an analogous relation holding when the outcome of the  first measurement  is $-1$.}
\end{quote}

Let us denote as $\Lambda^{(1)}({\bf a}_{\lambda}=+)$ the subset of $\Lambda$ such that, when only ${\bf A}^{(1)}$ is measured, one gets the outcome $+1$ (equivalently $\Lambda^{(1)}({\bf a}_{\lambda}=+)$ is the subset for which $P_{\lambda}({\bf a}_{\lambda}=+\vert*)=1)$. Similarly we define the subsets $\Lambda^{(1)}({\bf a}_{\lambda}=-)=\Lambda-\Lambda^{(1)}({\bf a}_{\lambda}=+), \Lambda^{(2)}({\bf b}_{\lambda}=+)$ and $\Lambda^{(2)}({\bf b}_{\lambda}=-)=\Lambda-\Lambda^{(2)}({\bf b}_{\lambda}=+)$. For our purposes it is also useful to define the four subsets of $\Lambda$, denoted collectively as $\Lambda^{(1,2)}({\bf a}_{\lambda}=i,{\bf b}_{\lambda}=j), i,j=\pm1$ which are  such that, for $\lambda\in \Lambda^{(1,2)}({\bf a}_{\lambda}=i,{\bf b}_{\lambda}=j)$, when both particles are subjected to the indicated measurement procedures yield the outcomes $i$ and $j$, respectively.

It is  obvious that {\bf Assumption L} implies:
\begin{eqnarray}
 \Lambda^{(1)}({\bf a}_{\lambda}=+) & \supseteq & \Lambda^{(1,2)}({\bf a}_{\lambda}=+,{\bf b}_{\lambda}=+)\cup \Lambda^{(1,2)}({\bf a}_{\lambda}=+,{\bf b}_{\lambda}=-),\nonumber \\
  \Lambda^{(1)}({\bf a}_{\lambda}=-) & \supseteq & \Lambda^{(1,2)}({\bf a}_{\lambda}=-,{\bf b}_{\lambda}=+)\cup \Lambda^{(1,2)}({\bf a}_{\lambda}=-,{\bf b}_{\lambda}=-).
  \end{eqnarray}
  
  \noindent  Since the two sets $ \Lambda^{(1)}({\bf a}_{\lambda}=i),\;i=\pm1,$  as well as the four sets $\Lambda^{(1,2)}({\bf a}_{\lambda}=i,{\bf b}_{\lambda}=j), \;\; i,j=\pm 1, $ are disjoint, and due to the fact that for any given $\lambda$ one gets precise outcomes in all  measurement processes, we must have:
  
  \begin{eqnarray}
 \Lambda^{(1)}({\bf a}_{\lambda}=+)\cup \Lambda^{(1)}({\bf a}_{\lambda}=-)& = &  \Lambda \nonumber \\
  \cup_{i,j}\Lambda^{(1,2)}({\bf a}_{\lambda}=i,{\bf b}_{\lambda}=j) & = & \Lambda.
 \end{eqnarray}
 
 \noindent These relations imply then that the equality sign must hold in Eq.(2.1):
 
 \begin{eqnarray}
 \Lambda^{(1)}({\bf a}_{\lambda}=+) & = & \Lambda^{(1,2)}({\bf a}_{\lambda}=+,{\bf b}_{\lambda}=+)\cup \Lambda^{(1,2)}({\bf a}_{\lambda}=+,{\bf b}_{\lambda}=-),\nonumber \\
  \Lambda^{(1)}({\bf a}_{\lambda}=-) &= & \Lambda^{(1,2)}({\bf a}_{\lambda}=-,{\bf b}_{\lambda}=+)\cup \Lambda^{(1,2)}({\bf a}_{\lambda}=-,{\bf b}_{\lambda}=-).
   \end{eqnarray}
 
 \noindent In exactly the same way one proves that:
 
 \begin{eqnarray}
 \Lambda^{(2)}({\bf b}_{\lambda}=+) & = & \Lambda^{(1,2)}({\bf a}_{\lambda}=+,{\bf b}_{\lambda}=+)\cup \Lambda^{(1,2)}({\bf a}_{\lambda}=-,{\bf b}_{\lambda}=+),\nonumber \\
  \Lambda^{(2)}({\bf b}_{\lambda}=-) &= & \Lambda^{(1,2)}({\bf a}_{\lambda}=+,{\bf b}_{\lambda}=-)\cup \Lambda^{(1,2)}({\bf a}_{\lambda}=-,{\bf b}_{\lambda}=-).
   \end{eqnarray}

\noindent From  Eqs.(2.3,4), taking into account that different outcomes are mutually exclusive,  we have: 
\begin{eqnarray}
\Lambda^{(1)}({\bf a}_{\lambda}=+)\cap\Lambda^{(2)}({\bf b}_{\lambda}=+)& \equiv & [\Lambda^{(1,2)}({\bf a}_{\lambda}=+,{\bf b}_{\lambda}=+)\cup \Lambda^{(1,2)}({\bf a}_{\lambda}=+,{\bf b}_{\lambda}=-)]\nonumber \\ & \cap & [\Lambda^{(1,2)}({\bf a}_{\lambda}=+,{\bf b}_{\lambda}=+)\cup \Lambda^{(1,2)}({\bf a}_{\lambda}=-,{\bf b}_{\lambda}=+)]\nonumber \\ & = &  \Lambda^{(1,2)}({\bf a}_{\lambda}=+,{\bf b}_{\lambda}=+).
\end{eqnarray}

\noindent Resorting  to the same argument for arbitrary choices of the indices $i,j$, we can conclude that {\bf Assumption L} implies:

\begin{equation}
 \Lambda^{(1,2)}({\bf a}_{\lambda}=i,{\bf b}_{\lambda}=j)=\Lambda^{(1)}({\bf a}_{\lambda}=i)\cap\Lambda^{(2)}({\bf b}_{\lambda}=j)\hspace {0.3 cm} \forall {\bf a,b},i,j.
 \end{equation}

 We define now a probability measure $\mu[\Gamma]$ on the subsets of the set $\Lambda$ according to:
 
 \begin{equation}
  \mu[\Gamma]=\int_{\Gamma}d\lambda\rho_{\psi}(\lambda).
  \end{equation}
  
  \noindent If, as already remarked, one takes into account that in a deterministic hidden variable theory all probabilities of the outcomes take the value 0 or 1,  the measures of sets like those considered before can be expressed as:
  
  \begin{eqnarray}
  \mu[\Lambda^{(1)}({\bf a}_{\lambda}=i)] & = & \int_{\Lambda}d\lambda\rho_{\psi}(\lambda)P_{\lambda}({\bf a}_{\lambda}=i|*)\nonumber \\
  \mu[\Lambda^{(1,2)}({\bf a}_{\lambda}=i,{\bf b}_{\lambda}=j)] & = & \int_{\Lambda}d\lambda\rho_{\psi}(\lambda)P_{\lambda}({\bf a}_{\lambda}=i,{\bf b}_{\lambda}=j),
  \end{eqnarray}
  
  \noindent     By taking the measures of the two sets at the left and right hand sides of equation (2.7) one then has:
  
  \begin{equation}
  \mu[ \Lambda^{(1,2)}({\bf a}_{\lambda}=i,{\bf b}_{\lambda}=j)]=\mu[\Lambda^{(1)}({\bf a}_{\lambda}=i)\cap\Lambda^{(2)}({\bf b}_{\lambda}=j)].
  \end{equation}
  
 \noindent On the other hand, we have
  
  \begin{equation}
 \mu[\Lambda^{(1)}({\bf a}_{\lambda}=i)\cap\Lambda^{(2)}({\bf b}_{\lambda}=j)]=\int_{\Lambda}d\lambda\rho_{\psi}(\lambda) P_{\lambda}({\bf a}_{\lambda}=i|*)\cdot P_{\lambda}(*|{\bf b}_{\lambda}=j).
  \end{equation}
  
 \noindent The validity of this equation is guaranteed by the fact that the integrand at the r.h.s. does not vanish (and takes the value 1) only when both probabilities appearing in it equal 1, which means, when $\lambda\in\Lambda^{(1)}({\bf a}_{\lambda}=i)$ and also $\lambda\in\Lambda^{(2)}({\bf b}_{\lambda}=j)$.
 
 Taking into account Eq.(2.9), one gets from the second of Eqs.(2.8) and Eq. (2.10) :
 
 \begin{equation}
 \int_{\Lambda}d\lambda\rho_{\psi}(\lambda)[P_{\lambda}({\bf a}_{\lambda}=i, {\bf b}_{\lambda}=j)-P_{\lambda} ({\bf a}_{\lambda}=i|*)\cdot P_{\lambda}(*|{\bf b}_{\lambda}=j)]=0.
 \end{equation}
 
\noindent Since $\rho_{\psi}(\lambda)$ is semipositive definite and the probabilities can take only the values 0 or 1,  the only negative contribution to the integral can come from a set $\Sigma$ such that, for $\lambda\in\Sigma$, $  
  P_{\lambda} ({\bf a}_{\lambda}=i|*)=P_{\lambda}(*|{\bf b}_{\lambda}=j)=1$ and $P_{\lambda}({\bf a}_{\lambda}=i, {\bf b}_{\lambda}=j)=0$. But such a set has measure zero as a consequence of our {\bf Assumption L}, because in it, while the measurement on one of the two constituents gives an outcome, it gives the opposite outcome when both particles are subjected to a measurement.
  
  The conclusion is then obvious: the vanishing of the integrand, i.e., the relation
  \begin{equation}
  P_{\lambda}({\bf a}_{\lambda}=i, {\bf b}_{\lambda}=j)=P_{\lambda} ({\bf a}_{\lambda}=i|*)\cdot P_{\lambda}(*|{\bf b}_{\lambda}=j),
  \end{equation}
  \noindent must hold almost everywhere. This relation asserts that the hidden variable theory under examination satisfies Bell's locality requirement\cite{Bell}, and, as such, when the state $\psi$ is entangled, cannot reproduce all predictions of quantum mechanics for such a state.
  
  The desired conclusion is then obtained: 
  
  \begin{quote}
  {\bf Conclusion}: {\it in any deterministic hidden variable theory which reproduces the predictions of quantum mechanics, in the case of  entangled states there are values of $\lambda$ and pairs of observables such that while if one measures only an observable of one of the particles   one gets the value $i$,  if one makes the same measurement on the considered particle but simultaneously one performs an appropriate measurement on the other particle, then the outcome of the measurement of the first particle turns out to be the opposite of the one obtained in the  case of only one measurement.The set of $\lambda's$ for which this happens has a non zero measure.}
  \end{quote}
  Some remarks:
  
  \begin{itemize}
  \item If we denote  as $\{B-Loc\}$ Bell's locality requirement (2.12) for the considered deterministic hidden variable theory, and as $\{QM\; pred\}$ the assumption that the considered theory reproduces the quantum predictions for all outcomes, the logic of the argument is quite straightforward: 
  \begin{equation}
 {\bf L} \Rightarrow \{B-Loc\}\Rightarrow \neg \{QM\; pred\}
  \end{equation}
  so that
   \begin{equation}
  \{QM\; pred\}\Rightarrow  \neg {\bf L} ,
  \end{equation}
  \item This result will appear as obvious to anybody who is familiar with quantum nonlocality. In fact it can be read as a proof of contextuality of any deterministic hidden variable theory reproducing the quantum predictions for an entangled state. However,  we consider it useful to have presented it in this way  since it shows explicitly that the outcome of an appropriate measurement of one particle  can be reversed by subjecting the other particle to an appropriate measurement. Obviously we have in mind the case in which  the two measurement processes take place and are completed in spacelike  regions, so that  our analysis is a proof of the {\it nonlocal character of contextuality}.  
  \item In particular in the quantum case of two far away particles in the singlet state we can claim that for any conceivable deterministic hidden variable theory predictively equivalent to quantum mechanics, there exists a non-empty subset $\Sigma$ of $\Lambda$ and a pair of directions ${\bf a}$ and ${\bf b}$, such that, for $\lambda\in\Sigma$, if one measures the spin component along ${\bf a}$ for particle 1 and does not perform any measurement on particle 2 he will get an outcome $i$, while, if two combined measurement processes along the considered directions are performed,  the outcome of the measurement on the first particle would turn out to be $-i$. In general, the two directions  ${\bf a}$ and ${\bf b}$ will turn out to be different. While we cannot exclude that the outcome of a measurement can be reversed by performing a measurement along the same direction on the partner particle, our analysis does not guarantee that this must happen for appropriate tests along the same direction. 
   \end{itemize}
\subsection{An alternative proof}   
   We have decided to derive our conclusion by following the procedure outlined in the previous section because it seems to us that it makes particularly clear the physical implications of quantum mechanics for any conceivable deterministic hidden variable theory equivalent to it. The same conclusion can be derived  by making reference to the proof \cite{Suppes,van Fraassen, Jarrett, Shimony}, holding in general, that Bell's locality assumption is equivalent to the conjunction of two other assumptions which have been denoted by A. Shimony as {\it Parameter Independence} and {\it Outcome Independence}, respectively. Let us fix our notation. We will continue to denote by $\lambda$ all parameters which provide, within the general theoretical scheme one is interested in, the complete specification of the state of an individual physical system. Within such a framework we consider  a standard EPR-Bohm like situation and we use the same symbols as before to denote the single and joint probability distributions of measurement outcomes. The only difference, at this stage, is that since we are not assuming determinism, such probabilities are no more constrained to have one of the values $\{0,1\}$. It is useful to introduce a further symbol  $P_{\lambda}({\bf a}_{\lambda}=i,{\bf b})=P_{\lambda}({\bf a}_{\lambda}=i,{\bf b}_{\lambda}=+)+P_{\lambda}({\bf a}_{\lambda}=i,{\bf b}_{\lambda}=-)$ expressing the probability of getting the outcome ${\bf a}_{\lambda}=i$ when two measurements are actually performed but one disregards the outcome obtained in one of them. In an analogous way one defines $P_{\lambda}({\bf a},{\bf b}_{\lambda}=j)$.
   
   In the above quoted papers it has been shown that Bell's locality condition Eq.(2.12) is equivalent to the conjunction of two logically  independent conditions:
   
   A. {\it Parameter Independence}:
   
   \begin{equation}
   P_{\lambda}({\bf a}_{\lambda}=i,{\bf b})=P_{\lambda}({\bf a}_{\lambda}=i\vert *),\;\;\;
   P_{\lambda}({\bf a},{\bf b}_{\lambda}=j)=P_{\lambda}(*\vert,{\bf b}_{\lambda}=j).
   \end{equation}
   
   B. {\it Outcome Independence}:
   
       \begin{equation}
   P_{\lambda}({\bf a}_{\lambda}=i,{\bf b}_{\lambda}=j)=P_{\lambda}({\bf a}_{\lambda}=i,{\bf b})\cdot
   P_{\lambda}({\bf a},{\bf b}_{\lambda}=j).
   \end{equation}  
   
   Note that the assumption of {\it Parameter Independence} amounts to claim that the probability of getting an outcome in one arm of  the apparatus is independent from the setting chosen at the other arm and coincides with the same probability when no measurement is performed there. The assumption of {\it Outcome Independence} implies that, when two measurements are actually performed,  the probability of an outcome at one wing does not depend on the outcome which is obtained at the other wing. We recall that quantum mechanics violates Bell's locality by violating only {\it Outcome Independence} , and, even more important for our purposes, that any {\it deterministic} theory  cannot violate  {\it Outcome Independence}. It follows that any deterministic theory equivalent to quantum mechanics must violate the {\it Parameter Independence} requirement. This in turn implies, from (2.15), that there must be $\lambda$'s, ${\bf a}$'s and ${\bf b}$'s, for which, while if one performs a measurement on one of the particles one gets an outcome, one gets the opposite outcome when two measurements are performed. 
    
    This is another, more formal way of showing that any deterministic hidden variable theory agreeing with quantum mechanics  turns out to be unavoidably {\it non-locally contextual} in the very precise sense we have  made clear.  However, due to the fact that the violation of {\it Parameter Independence} might occur only for different directions {\bf a} and {\bf b}, our result is not so strong as we need to go on with our argument. Let us then pass to consider more articulated situations.
    
\section{A more useful analysis}

To refine our analysis we recall that in the case of an Hilbert space of dimensionality equal or greater than 3, the fact that any deterministic hidden variable theory   must exhibit a contextual nature is a consequence of a general theorem due to Gleason \cite {Gleason}. The  argument goes as follows: one considers a system whose Hilbert space is $N$-dimensional  and a complete set  $\{P_{i}\}$ ($\sum_{i=1}^{N}P_{i}=1)$  of commuting orthogonal projection operators  on one dimensional manifolds . Since the eigenvalues of these operators are only 0,1, and since the sum of their values must be 1, any deterministic hidden variable theory, given any set of $N$ orthogonal unit vectors, must assign the value zero to all of them but one. This is  impossible, as it has been explicitly proved for the first time\footnote{It is worth mentioning that, as stated above, the conclusion drawn by Bell is a consequence of the Gleason Theorem. However Gleason argument was addressed to generalised probability measures on the lattice of projection operators, while it was Bell to derive in a simple way the precise result which is relevant for the problem of  contextuality.} by J.S. Bell \cite{Bell1}, for $N\ge 3$. 

In the case of a single particle of spin 1 Kochen and Specker\cite{Kochen} have considered an analogous but slightly different problem. For such a system the squares of the spin-components along any 3 orthogonal directions commute among themselves. Since the eigenvalues of the just considered operators are $0$ and $1$ and due to the fact that the sum of the squares equals 2 for any orthogonal triple  {\bf i}, {\bf j} and {\bf k} of unit vectors in ordinary space, any deterministic hidden variable theory must assign to two of the squares of the spin components the value 1 and to the remaining one the value 0.  Kochen and Specker themselves have  identified a precise set of 117 directions for which this requirement cannot be satisfied. This implies that at least one of the squares of the spin components  has a contextual character, in the precise sense that, the hidden variable being assigned, it takes either the value 0 or 1 according to which one of the commuting squares of the components along an appropriate direction orthogonal to the first direction is measured. 

Subsequently the number of directions leading to a contradiction has been reduced\cite{Peres} to 33 by A. Peres. Conway and Kochen \cite{Conway} have recently called a ``101-function" a function on the set of directions such that its values on each orthogonal triple would be 1,0,1 in some order. It is a consequence of the beautiful and elegant theorem proved by Kochen and Specker  that: {\it there is no 101-function for the $\pm 33$ directions identified by Peres}.
 
 For the case of 2 spin-$1/2$ particles,  Peres himself \cite{Peres} has shown that actually one can identify 24 rays  of the 4-dimensional Hilbert space such that the value of one of the projection operators along them must have a contextual value. We also stress that the problem of the nonlocal contextuality associated to entangled states of two far away spin-1 particles has recently attracted a lot of attention since its consideration  is the starting point of two quite interesting papers by Conway and Kochen \cite{Conway, Conway3} in which the so called {\it Free will theorem} has been derived\footnote{For a debate of the important questions raised by the just quoted papers we refer the reader to refs.\cite{Adler, Ghirardi, Tumulka, Conway2,Hooft}}.

Let us consider  the rotationally invariant entangled state of two spin 1 particles which has been used by Conway and Kochen for their argument:

\begin{equation}
\vert\Psi(1,2)\rangle=\frac{1}{\sqrt{3}}[\vert 1\rangle_{1}\otimes\vert -1\rangle_{2}-\vert 0\rangle_{1}\otimes\vert 0\rangle_{2}+\vert -1\rangle_{1}\otimes\vert 1\rangle_{2}].
\end{equation}

\noindent Some important remarks concerning this state are at order:
\begin{itemize}

\item The state is an eigenstate of the square of the total spin  $[{\bf S}^{(1,2)}]^{2}=[{\bf S}^{(1)}+{\bf S}^{(1)}]^{2}$ of the composite system pertaining to the eigenvalue $0$:
\begin{equation}
[{\bf S}^{(1,2)}]^{2}\vert \Psi(1,2)\rangle=0,
\end{equation}
\noindent and, as such, it is rotationally invariant. This means that, in Eq, (3.1) one can choose in an arbitrary way the direction along which the projections of the spins take the indicated values 1,0,-1.

\item As it follows trivially from the fact that $[S^{(i)}_{x}]^{2}+[S^{(i)}_{y}]^{2}+[S^{(1i)}_{z}]^{2}=2\cdot 1^{(i)}$, our state $\vert\Psi (1,2)\rangle$ (as well as any conceivable vector of the Hilbert space ${\mathcal H}^{(1}\otimes {\mathcal H}^{(2)}$) is an eigenstate belonging to the eigenvalue 2 (in units of $\hbar^{2}$) of the sum of the squares of the components of the spin of any one of the component subsystems along the three orthogonal directions (${\bf x},{\bf y},{\bf z}$) :
\begin{eqnarray}
\{[S^{(1)}_{x}]^{2}  \otimes  1^{(2) } & + & [S^{(1)}_{y}]^{(2)}\otimes 1^{2 }+[S^{(1)}_{z}]^{(2)}\otimes 1^{2 }\}\vert \Psi\rangle=2\vert\Psi\rangle \nonumber \\
\{1^{(1)}\otimes [S^{(2)}_{x}]^{2}   & + & 1^{(1)}\otimes [S^{(2)}_{y}]^{2}+1^{(1)}\otimes [S^{(2)}_{z}]^{2}\}\vert \Psi\rangle=2\vert\Psi\rangle
\end{eqnarray}

\item An interesting and non trivial  fact is that, for any of the three terms within the curly brackets at the l.h.s. of the above equations (3.3), one can change  the type of particle to which the component refers, without changing the fact that the state remains an eigenstate belonging to the same eigenvalue\footnote{This is obviously related to the fact that, for the considered state, the outcomes of spin measurements along the same direction for the two particles give always perfectly correlated results.}. For example,  if one replaces  in the second of Eqs.(3.3) the {\it z}-component of particle 2 with the one of particle 1 one still has:

\begin{equation}
\{1^{(1)}\otimes [S^{(2)}_{x}]^{2}    +  1^{(1)}\otimes [S^{(2)}_{y}]^{2}+[S^{(1)}_{z}]^{2}\otimes1^{(2)} \}\vert \Psi\rangle=2\vert\Psi\rangle
\end{equation}

\item The validity of Eq. (3.4) as well as the one of all the other similar ones obtained by switching from particle 1 to particle 2 the various terms of equations (3.3), is easily proved. A straightforward way to realize this fact derives from rewriting, by making use of the fact that $ [S^{(2)}_{x}]^{2}+ [S^{(2)}_{y}]^{2}=2\cdot 1^{(2)} - [S^{(2)}_{z}]^{2}$,          the operator in (3.4)  in the following way:
\begin{eqnarray}
\{1^{(1)}\otimes [S^{(2)}_{x}]^{2}   &  + & 1^{(1)}\otimes [S^{(2)}_{y}]^{2}+[S^{(1)}_{z}]^{2}\otimes1^{(2)} \} \nonumber \\=\{2\cdot 1^{(1)}\otimes1^{(2)}   &  - & 1^{(1)}\otimes [S^{(2)}_{z}]^{2}+[S^{(1)}_{z}]^{2}\otimes1^{(2)} \},
\end{eqnarray}
\noindent from which one immediately realizes that all three states $\vert 1\rangle_{1}\otimes\vert -1\rangle_{2}, \vert -1\rangle_{1}\otimes\vert 1\rangle_{2}$ and $\vert 0\rangle_{1}\otimes\vert 0\rangle_{2}$ appearing in the state (3.1) (as well as the two other states $\vert 1\rangle_{1}\otimes\vert 1\rangle_{2}$ and $\vert -1\rangle_{1}\otimes\vert -1\rangle_{2}$) are eigenvectors of this operator belonging to the eigenvalue 2. 

\item It has to be stressed that  the just mentioned property of our state (3.1) of being an eigenvector of the operator (3.5) associated to the considered eigenvalue depends crucially on the specific form of the state itself.  In fact the states $\vert 1\rangle_{1}\otimes\vert 0\rangle_{2}$ and $\vert -1\rangle_{1}\otimes\vert 0\rangle_{2}$ are eigenstates of such an operator belonging to the eigenvalue 3, while the states    $\vert 0\rangle_{1}\otimes\vert 1\rangle_{2}$ and $\vert 0\rangle_{1}\otimes\vert -1\rangle_{2}$ belong to the eigenvalue 1. This shows once more that, while Eqs. (3.3), as already remarked, hold trivially for any state  of the composite system, Eq. (3.4) has a precise physical meaning.
\end{itemize}

We prove now the following 
\begin{quote}
\noindent  Theorem: {\it For any arbitrary triple of orthogonal directions (${\bf i}, {\bf j}, {\bf k}$), one has}\footnote{It goes without saying that in Eq. (3.6), one can change the particle labels ($^{(1)},^{(2)}$) attached to each of the indices ($i,j,k$)   as he likes.} :

\begin{equation}
\{1^{(1)}\otimes [S^{(2)}_{i}]^{2}    +  1^{(1)}\otimes [S^{(2)}_{j}]^{2}+[S^{(1)}_{k}]^{2}\otimes1^{(2)} \}\vert \Psi\rangle=2\vert\Psi\rangle
\end{equation}

\end{quote}

Proof: one   considers Eq. (3.4) and takes into account  the rotation, characterized by the vector ${\bf \omega}$, which leads from the orthogonal triple (${\bf x},{\bf y},{\bf z}$) to the orthogonal triple (${\bf i},{\bf j},{\bf k}$). Such a rotation is implemented by the unitary operator of the space of the two particles $U_{{\bf \omega}}(1,2)=e^{-i{\bf \omega}\cdot {\bf S}}\equiv e^{-i{\bf \omega}\cdot ({\bf S}^{1}+{\bf S}^{(2)})} $. The operator $U_{{\bf \omega}}$  rotates the single particle spin components in such a way to transform those of Eq.(3.4) in those appearing in Eq. (3.6).  In formulae:

\begin{eqnarray}
U(1,2)\{1^{(1)} & \otimes &  [S^{(2)}_{x}]^{2}    +  1^{(1)}\otimes [S^{(2)}_{y}]^{2}+[S^{(1)}_{z}]^{2}\otimes1^{(2)} \}U^{\dagger}(1,2)U(1,2)\vert\Psi\rangle = \nonumber \\
=\{1^{(1)}  \otimes  [S^{(2)}_{i}]^{2}  &  + & 1^{(1)}\otimes [S^{(2)}_{j}]^{2}+[S^{(1)}_{k}]^{2}\otimes1^{(2)} \}U(1,2)\vert \Psi\rangle =2U(1,2)\vert\Psi\rangle.
\end{eqnarray}
\noindent Taking into account the rotational invariance of our state $U(1,2)\vert\Psi\rangle=\vert\Psi\rangle$, one gets eq. (3.6)\rule{3mm}{3mm}

Now our game is almost over. In any deterministic hidden variable theory which reproduces the predictions of quantum mechanics for the state (3.1), the fact that all the terms at the l.h.s. of Eq. (3.6) can take only the value $0$ or $1$, and the fact that the equation itself implies that their sum must equal 2, leads us to the same problem we have met for a single particle of spin 1. In fact, any theory of the kind we are envisaging requires the identification of  a 101-function on all orthogonal triples of axes in real three-dimensional space. If consideration is given to   the 33 directions identified by  Peres  we can take into account the  proof of Kochen and Specker that there is no 101-function for them. An argument which is perfectly analogous to the one leading to the contextuality in the case of one particle leads to the same conclusion of the unavoidable nonlocal contextuality in our case\footnote{For an appropriate understanding of the conceptual difference between the two cases we invite the reader to take into account that here an essential reference is made to the state we have assumed to describe our composite system. Our conclusion, which differs from those found in the literature, does not hold for an arbitrary state (as it is the case for a single particle) but for the specific maximally entangled state (3.1),  just for its property of being  rotationally invariant.}. 

In fact, the non-existence of a 101-function for the 33 considered directions implies, as before, there  must be directions ${\bf k}$ such that the fact that one gets the outcome $0$ or $1$ in a mesurement of $[S^{(1)}_{k}]^{2}$, is not uniquely determined  by the value  $\lambda\in\Lambda$ of the hidden variables  but it depends also from the choice of the two orthogonal directions along which a far away observer measures the square of the spin components of the other particle. Said differently, there  exists a non empty subset $\Omega$ of $\Lambda$, a direction ${\bf k}$ and two pairs of directions (${\bf i}$ , ${\bf j}$) and (${\bf r}$ , ${\bf s}$) such that, for $\lambda\in\Omega$, while if one measures   $[S_{\bf k}^{(1)}]^{2},[S_{\bf i}^{(2)}]^{2},[S_{\bf j}^{(2)}]^{2}$ one gets the three outcomes (0,1,1), in the case one measures  $[S_{\bf k}^{(1)}]^{2},[S_{\bf r}^{(2)}]^{2},[S_{\bf s}^{(2)}]^{2}$ one gets the outcomes (1,0,1) or (1,1,0). Since the hidden variable  theory, once $\lambda$ is fixed, attaches a precise value (either 0 or 1) to the outcome one  gets if the only measurement which is performed is the one of $[S_{\bf k}^{(1)}]^{2}$, the above analysis implies that for a nonempty subset $\Sigma$ of $\Lambda$, such an outcome must  change (i.e. it becomes 1 or 0, respectively) when an appropriate pair of measurements  are performed on the second, far away particle.

The conclusion is analogous to the one of the previous analysis:
 \begin{quote}
  {\bf Conclusion}: {\it in any deterministic hidden variable theory which reproduces the predictions of quantum mechanics for the rotationally invariant entangled state (3.1) of two spin 1 particles there are values of $\lambda$ and pairs of directions {\bf k} and {\bf i}, (or, alternatively {\bf k} and ${\bf k\times i}$) such that,  for them, while if one measures only the square of the spin component  of one of the particles  along {\bf k} one gets the value $0$,  if one makes the same measurement on the considered particle but simultaneously one performs an appropriate measurement  along one of the two indicated directions for the other particle, then the outcome of the measurement of the "first" particle turns out to change from 0 to 1 and the outcome for the second particle is zero.The set of $\lambda's$ for which this happens has a non zero measure.}
  \end{quote}

We can now pass to perform a critical analysis of the use of counterfactual arguments involving  entangled states of far away composite systems within the considered scenario.

\section{Contextuality and counterfactual arguments}
In this section we will make use of the elementary derivation of the previous section to reconsider the problem of the alleged {\it spooky action at-a-distance} occurring in an EPR-like situation. This instantaneous action is usually related to the fact that, in a situation of the kind envisaged by Einstein Podolsky and Rosen\cite{Einstein} as reformulated by D. Bohm \cite{Bohm}, when an observer performs a measurement on one of the two far away constituents of a composite system he can infer, from his knowledge of the outcome, the precise quantum state of the far away constituent, and, accordingly, he can ``predict with certainty" the outcome of a prospective measurement on the other constituent.
	It goes without saying that the line of thought we are considering is based on a counterfactual argument. In fact the spooky action at a distance follows from the fact that one considers legitimate to make a statement of the sort ``if an observer would perform a measurement on the far away constituent he will get such and such an outcome", the basic premise which makes legitimate, according to EPR, the claim that the system possesses  objectively the element of physical reality associate to the certain outcome. Then, if compleness of the theory is assumed, such an objective property can be claimed to have  been {\it produced} by the far away measurement process; it is precisely on the basis of this and of the assumption that no istantaneous change of the property of a system can be produced by a space-like ac tion, that EPR have been led to deny that quantum mechanics is a complete theory.
		
		At this point we pass to analyze the structure of that form of common-sense non-monotonic inference which is referred to as counterfactual reasoning. A counterfactual is a statement of the "if ... then " type in which the antecedent is known (or expected) to be false. In our discussion it will be assumed to be false. As a consequence, on the na\"{i}ve formalization using $\Rightarrow$ for "if ... then", all counterfactual arguments would be true, contrary to our intuitive understanding of them. 
		
		A widely accepted way of dealing with counterfactuals is the one discussed in the lucid text {\it Counterfactuals}, by D. Lewis, who treats them\cite{Lewis} as variable strict conditionals in the usual possible worlds semantic for modal logic. This requires to make precise how the truth value at a given possible world of a counterfactual depends on the truth values at various possible worlds of its antecedent and consequent.
		
		Let us denote the counterfactual ``{\it if} $\phi$ were true, {\it then} $\psi$ would be true" as ``$\phi \Box\rightarrow \psi$", for propositions $\phi$ and $\psi$. Then one proposes the following truth condition:
		\begin{quote}
		$\phi \Box \rightarrow \psi$ is true at word {\it w} if {\it either} (i) there are no possible worlds in which $\phi$ is true, {\it or} (ii) some worlds where both $\phi$ and $\psi$ are true are more similar (closer) to {\it w} than any world in which $\phi$ is true and $\psi$ is false. 
		
		Obviously, one has to specify in a precise way the possible worlds one is taking into account; this is achieved by assigning to the actual world {\it w} a set of worlds $S_{w}$ called the sphere of accessibility around {\it w}.
		\end{quote}
		
		We agree with Lewis that the concept of similarity between worlds is to some extent problematic, but it is just the necessity of making precise this point which will give rise to the puzzling situation we are going to face.
		
		Let us assume  that we are in a situation like the one devised above: we have two far away spin 1 particles in the  state (3.1) and one observer, let us call him Alice, who performs a  measurement of the square of the spin  along the {\bf k}-axis on her particle. Suppose she gets the outcome $0$, and let us consider the counterfactual statement ``if Bob (the far away observer) would make a spin measurement of the spin component of his particle along any direction  orthogonal to {\bf k}  he would certainly get the outcome $1$". Note that the logic underlying this statement is strictly analogous to the one used in the EPR situation. In fact, according to quantum mechanics, for the considered outcome the state (3.1) will instantaneously be reduced to the state $\vert 0^{\diamond}\rangle_{1}\otimes\vert 0^{\diamond}\rangle_{2}$, where we have used the apex $\diamond$ to denote the eigenstates of the observables $[S_{k}^{(i)}],\;(i=1,2)$, pertaining to the eigenvalue 0. According to quantum mechanics, for such a state a measurement of $[S_{l}^{(2)}]^{2}$ on particle 2 along any direction ${\bf l}$ orthogonal to ${\bf k}$ would give the outcome 1: Alice can {\it predict with certainity} the outcome of a prospective measurement.

		With these premises we can now take into account the  nonlocal process occurring as a consequence of Alice's measurement, and analyze the counterfactual statement we have presented above. Let us perform the possible worlds analysis by having in mind, in place of standard quantum mechanics,  a deterministic hidden variables nonlocal model of the kind we have considered in the previous section. 
		
		Let us  denote as $\Lambda^{(1)}_{0,{\bf k}}$ the subset of $\Lambda$ for which, when only the measurement of the square of the component of particle 1 along the direction {\bf k} is performed one gets the outcome 0, and by  $\Lambda^{(1,2)}_{1,{\bf k};0,{\bf l}}$ the subset of $\Lambda$ such that, when two measurements are performed, the one on particle 1 along {\bf k}  and the one on the second particle along {\bf l},  one gets the ordered pair of outcomes (1,0).  Then the  conclusion  of the previous section can be expressed in formal terms in the following way:
	\begin{equation}
 \exists {\bf k},{\bf l}\;\vert\;\Lambda^{(1)}_{0,{\bf k}}\cap\Lambda^{(1,2)}_{1,{\bf k};0,{\bf l}}\neq\emptyset .
\end{equation}
\noindent Accordingly, if we denote as $\Sigma$ the set defined by (4.1), we can state that, for $\lambda\in\Sigma$ the outcome that Alice has got in her measurement process, i.e. 0, changes to  1 when two measurement are performed, the same by Alice and one  along {\bf l} by Bob who gets the outcome  0.

Now we come back to our counterfactual statement. If we denote as $\phi$ the proposition ``Bob performs a measurement along {\bf l}" and as $ \psi$ the statement ``Bob gets the outcome 1", then the counterfactual statement 
$\phi \Box \rightarrow \psi$ cannot be made. Let us analyze better the situation. In the actual world, only Alice performs a measurement and she does not know, and cannot know, the  value of the hidden variables in the actual world. Now we are compelled to identify the sphere of accessibility from the actual world. Usually, in physics, the criterion to identify the accessible worlds rests on the so called ``inevitability at the time {\it t} at which one makes the statement". This position requires to consider as nearer to the actual world the worlds which coincide with the actual one at times previous to {\it t} and in which all physical laws  of the actual world hold.

In our case we are in troubles because at time {\it t} both the value of the hidden variable as well as the outcome obtained by Alice are given and should be kept unchanged according to the previous criterion. The situation can be summarized as follows:

\begin{itemize}
\item If one chooses as alternative worlds those in which the hidden variable $\lambda$ takes the same value as the one it has in the actual world (which amounts to keep fixed the ``state" characterizing the system) and if the actual $\lambda$ belongs to $\Sigma$, the statement $\phi \Box \rightarrow \psi$ turns out to be false because actually $\phi\Rightarrow\neg\psi$ is true.
\item If one  chooses as alternative (closer) to the actual world the worlds in which the outcome 0 obtained by Alice in the actual world is kept fixed then the implication  $\phi\Rightarrow\psi$ is correct. However we cannot ignore that, in so doing, we are including among the accessible worlds also worlds for which if in place of two measurements only the one performed by Alice would take place, one would obtain the outcome 1 instead of the one obtained by Alice. In fact, since $\Lambda^{(1,2)}_{0,{\bf k};1,{\bf l}}$ cannot be entirely contained in $\Lambda^{(1)}_{0,{\bf k}}$ it unavoidably has a non empty intersection with  $\Lambda^{(1)}_{1,{\bf k}}$ (the complement of $\Lambda^{(1)}_{0,{\bf k}}$) a fact which implies the odd conclusion we have just drawn.
\end{itemize}

It seems obvious to us that both alternatives lead to an unacceptable situation, a conclusion which simply means that in a  nonlocal context one cannot make claims on what would have  occurred at a space-like separation if the situation there would be different from the one of the actual world. This conclusion leads to assert that claims of the sort we have considered are not legitimate in the considered context and, accordingly, that it is unappropriate to make reference to {\it spooky actions at-a-distance}.

A perfectly similar argument could be developed also for joint measurements involving two spin-1/2 particles in the singlet state when one deals with a nonlocal deterministic hidden variable theory in which one can  reverse an  outcome also by resorting to a measurement on the partner particle  along the same direction. In fact in such a case, on the one side the existence of perfect (anti)correlations allows to infer (with probability equal to 1) from the outcome obtained by Alice the outcome that would be obtained by Bob would he perform a measurement, while, on the other, the deterministic hidden variable theory would give rise to the same problems we have discussed here for the entangled state of two spin-1 particles. However, we have not been able to prove that all deterministic hidden variable theories equivalent to quantum mechanics in the case of two spin 1/2 particles must imply the just mentioned reversal of the outcome when one passes from the case of only one  to the case of two measurements, if both take place along the same direction. This is why we have performed the more conclusive analysis of Section 3.

We conclude this paper by mentioning that, several years ago, we have developed
a similar analysis within the conceptual framework of standard quantum mechanics and we have reached the conclusion that in a relativistic and nonlocal context the use of counterfactuals concerning events which are spacelike separated might lead to illegitimate claims. We refer the reader to ref.\cite{ghirardi} for a discussion of this problem.

\section*{Acknowledgment}
We thank Prof. G. Calucci for useful and illuminating discussions.

\end{document}